\documentclass[prl,twocolumn,superscriptaddress,showpacs,amsmath,amssymb,floatfix]{revtex4}
\usepackage{graphicx,color,epsfig,bm,float}% Include figure files

\begin{document}

\title{Operator folding and matrix product states in linearly-coupled bosonic arrays}

\author{Jose Reslen}

\affiliation{Coordinaci\'on de F\'{\i}sica, Universidad del Atl\'antico, Km. 7 Antigua v\'{\i}a a Puerto Colombia, A.A. 1890, Barranquilla, Colombia.}

\date{\today}

\begin{abstract}
A protocol to obtain the matrix product state representation of 
a class of boson states is introduced. The proposal is presented
in the context of linear systems and is tested by performing
simulations of a reference model. The method can be applied
regardless of the details of the coupling among modes and can 
be used to extract the most significant contribution of the 
tensorial representation. Characteristic issues 
as well as potential variants of the proposed protocol are discussed.
\end{abstract}

\maketitle

\section{Introduction}

The realization that  quantum states can be written in terms
of a tensor network whose elements display interesting properties has
prompted a wealth of research in what is nowadays known as the
field of Matrix Product State (MPS) \cite{Ulrich}. 
Although the properties of MPS can be exploited in 
a variety of ways, it is Time Evolving Block Decimation (TEBD) 
and Density Matrix Renormalization Group (DMRG), together with 
their variants, that have proved highly robust and appropriate 
in most situations of interest. However, other methods have also 
been proposed, for example, in the area of infinite chains,
where the calculation of local mean values can be formulated
in terms of bundled tensor networks, or in the area of
Gaussian states, where the MPS network is obtained as projections of 
highly entangled states \cite{Cirac1}. MPS offers a view
that is particularly convenient in variational approaches,
where some physical state is obtained by renormalizing a
tensor network. This has led to an interest
in classes of states that can be efficiently simulated \cite{Evenbly}.
Notwithstanding its  recurrent use in spin models, the relevance of MPS is 
especially notorious in bosonic systems. In this context, 
the application of TEBD has allowed the numerical exploration 
of boson chains under different conditions \cite{Reslen2,Danshita,Muth,Lacki,Hu}
revealing phases and regimes with very interesting properties.

Perhaps the most elementary way of representing a quantum
state is as a set of complex coefficients derived by
writing such a state as a superposition of elements of a basis.
In what respects to indistinguishable particles, the
basis is constituted by occupation states upon which
ladder operators can raise or lower the associated number of particles.
Because any of these states can be put in terms of ladder operators
acting on the vacuum, it is possible to envisage a representation
relative to such operators. 
This approach is practical, for example, when the symmetries
of the problem allow an advantageous handling of the
Heisenberg equations \cite{Prior}. This is seen in linear systems
where the underlying physics is driven by 
interference and single body (SB) effects. These systems
are quite recurrent, not only as realistic descriptions
of physical phenomena, such as optical fields \cite{Buzek} 
or weakly-interacting Bose-Einstein condensates, but also as 
modeling tools. The latter case is manifest, for
instance, in the framework of the mean field or Hartree-Fock approximation. 
Insight in this direction must therefore be of significance 

In the development that follows, a method is proposed to go from a 
representation of a bosonic state in terms of operators to a canonical MPS representation. 
The analysis makes use of the properties of both representations
and the central argument does not involve approximations. Results 
obtained using the proposed technique are compared against benchmark data. 
It is pointed out that the range of applicability does not depend
on boundary conditions or number of next-neighbors, but rather
on whether the state can be put in a compatible form.
In the final part, potential applications and complementary
remarks are set forth.

\section{Linear bosonic systems}

Following a second quantization scheme, let us propose a system of $M$ bosons. 
Every boson can occupy $N$ quantum
levels which are characterized by the bosonic operators $\hat{a}_j$
and $\hat{a}_k^{\dagger}$ satisfying $[\hat{a}_j,\hat{a}_k^{\dagger}]=\delta_j^k$ 
and $[\hat{a}_j,\hat{a}_k]=0$  with $j,k=1,2,...,N$. In absence of interaction, 
the Hamiltonian can be written as

\begin{equation}
\hat{H} = \sum_{j=1}^N\sum_{k=1}^N h_{j,k} \hat{a}_j^{\dagger} \hat{a}_k, \text{ } h_{j,k} = h_{k,j}^*. 
\label{eq:1}
\end{equation}

Matrix $h_{j,k}$ ($\hat{h}$) is the Hamiltonian when $M=1$. $\hat{h}$ also 
defines the operator dynamics according to

\begin{equation}
\frac{d \hat {\alpha}_j^{\dagger}}{d t} = -i \sum_{k=1}^N h_{j,k} \hat {\alpha}_k^{\dagger}, 
\label{eq:3}
\end{equation}

which can be obtained by differentiation of $\hat{\alpha}_j^{\dagger} = e^{-it\hat{H}} \hat{a}_j^{\dagger} e^{it\hat{H}}$ ($\hbar=1$).
A product of local Fock states $|n_1,n_2,...,n_{N}\rangle$, for which $n_1+n_2+...+n_{N}=M$, evolves as

\begin{equation}
|\psi(t) \rangle = \prod_{q=1}^{N} \frac{\left( \hat{\alpha}_q^{\dagger} \right)^{n_q}}{\sqrt{n_q!}}|0\rangle,	
\label{eq:4}
\end{equation}

where $|0\rangle$ is the state with no bosons. More complex configurations can be
constructed as superposition of these states. Now let $\epsilon_{l}$ 
be an eigenvalue of $\hat{h}$ corresponding to the normalized eigenstate $|\epsilon_l\rangle$

\begin{equation}
\sum_{k=1}^N h_{j,k} \epsilon_{k,l} = \epsilon_{l} \epsilon_{j,l} \text{ } (l,j=1,2,...,N).
\label{eq:5}
\end{equation}

An eigenstate $|E_{n_1...n_{N}}\rangle$ of $\hat{H}$ with 
eigenenergy $E_{n_1...n_{N}} = n_1 \epsilon_1 + n_2 \epsilon_2 + ... + n_{N} \epsilon_{N}$
can be built as a product of SB eigenmodes as 

\begin{equation}
|E_{n_1...n_{N}}\rangle = \prod_{q=1}^N \frac{1}{\sqrt{n_q!}} \left( \sum_{j=1}^N \epsilon_{j,q} \hat{a}_{j}^{\dagger} \right)^{n_q} |0\rangle.
\label{eq:2}
\end{equation}

The size of the basis is $(N+M-1)!/M!(N-1)!$. It can be seen that the
state of a system of free bosons is determined fundamentally by the contribution of the SB
Hamiltonian and the interference effects arising from indistinguishability, which is 
implicit in the bosonic operators. This characteristic renders the system into a linear
regime, where a composition of solutions of $\hat {H}$, like in Eq. (\ref{eq:2}),
is also a solution, and a SB eigenmode remains physically unaffected by other SB eigenmodes.

\section{Bosonic states in MPS form}

In order to establish a ground to perform the transition to MPS, let us imagine that bosons 
are arranged in a chain with open boundary conditions. This assumption however does not 
need to coincide with the real boundary conditions of the problem. A site in the chain is labeled
by the integer $n$ ranging from $1$ in the right end to $N$ in the left end. Using MPS, the quantum state 
can be represented as a superposition of non-local states
in the following way (up to few changes, the notation in \cite{Vidal} is followed)

\begin{equation}
|\psi \rangle = \sum_{\mu \nu p} \lambda_\nu^{[n]} \Gamma_{\nu,\mu}^{[n]} (p) \lambda_\mu^{[n-1]} |\nu^{[N:n+1]} \rangle |p^{[n]} \rangle |\mu^{[n-1:1]}  \rangle  .
\label{eq:6}
\end{equation}

$|\mu^{[n-1:1]}\rangle$ and $|\nu^{[N:n+1]}\rangle$ are, in that order, Schmidt vectors to the right and left of site $n$
(superscripts indicate the vector subspace). Notice that on each case such Schmidt vectors 
belong to different decompositions of the chain. $\lambda_\mu^{[n-1]}$ and $\lambda_\nu^{[n]}$ are the
Schmidt coefficients associated to such decompositions. The states $|p^{[n]}\rangle$ are elements
of a local basis at site $n$. For bosons, it is convenient to choose a local Fock basis. The complex coefficient
$\Gamma_{\nu,\mu}^{[n]} (p)$ determines the contribution of a basis state to the superposition. 
Integer $p$ is an occupation number and ranges from $0$ to $M$. Integers $\mu$ and $\nu$ are  
labels of two distinct sets of Schmidt vectors. The maximum number of these vectors over all possible
bipartite decompositions of the chain is called $\chi$. An important aspect of the 
MPS representation is that by adjusting $\chi$ it is possible to control the number of coefficients
employed to describe the state. This allows to approximate huge states by
retaining the most significant contribution of their respective MPS representations
(the part linked to the biggest $\lambda$s). The
set of tensors $\{ \Gamma_{\nu,\mu}^{[n]} (p),\lambda_\mu^{[n]}. \forall (\mu,\nu,p,n) \}$  
is a representation of $|\psi\rangle$ that can be updated when an unitary transformation
is applied on a pair of consecutive sites.
In what follows, it is shown how this feature can be applied to put states like
(\ref{eq:4}) or (\ref{eq:2}) in MPS form. 

Let us start by considering the simplified case where $n_1$ bosons occupy the same 
arbitrary SB state. The state can then be written in terms of a non-diagonal mode (NDM) as

\begin{equation}
|\psi \rangle = \frac{1}{\sqrt{n_1!}} \left( c_{1,1} \hat{a}_1^{\dagger} + c_{2,1} \hat{a}_2^{\dagger} + ... + c_{N,1} \hat{a}_N^{\dagger} \right )^{n_1} |0\rangle.
\label{eq:7}
\end{equation}

The meaning of the second subscript in the coefficients is explained further down. 
Normalization of $|\psi\rangle$ requires

\begin{equation}
\sum_{j=1}^N |c_{j,1}|^2 = 1.
\end{equation}

In a first step all these coefficients are to be made real. The idea is to operate on $|\psi\rangle$ with a series of 
local unitary transformations that act on the operators and take away the complex
phases of the coefficients as follows 

\begin{equation}
e^{-i \phi_{l,1} \hat{a}_l^{\dagger} \hat{a}_l} \hat{a}_l^{\dagger} e^{i \phi_{l,1} \hat{a}_l^{\dagger} \hat{a}_l} = e^{-i \phi_{l,1}} \hat{a}_l^{\dagger} \Rightarrow c_{l,1} \rightarrow |c_{l,1}|,
\label{eq:8}
\end{equation}

where $\phi_{l,1}$ is the phase of $c_{l,1}$. This is done for $l=1,2,...,N$. 
The order in which the transformations are applied is not important. Next, 
a rotation operation is applied on a couple of neighbor sites using 
the angular momentum operator 

\begin{equation}
\hat{J}_{j+1,j}^y = \frac{1}{2i} \left( \hat{a}_{j+1}^{\dagger} \hat{a}_{j} - \hat{a}_{j}^{\dagger} \hat{a}_{j+1} \right).
\label{eq:10}
\end{equation}

Explicitly, this transformation reads,

\begin{eqnarray}
e^{-i \theta_{j,1} \hat{J}_{j+1,j}^y} \left (  |c_{j+1,1}| \hat{a}_{j+1}^{\dagger} +  |c_{j,1}| \hat{a}_{j}^{\dagger} \right )    e^{i \theta_{j,1} \hat{J}_{j+1,j}^y} \nonumber \\
= \left ( |c_{j+1,1}| \cos \left( \frac{\theta_{j,1}}{2} \right) - |c_{j,1}| \sin \left( \frac{\theta_{j,1}}{2} \right)  \right )\hat{a}_{j+1}^{\dagger} \nonumber \\
+ \left ( |c_{j+1,1}| \sin \left( \frac{\theta_{j,1}}{2} \right) + |c_{j,1}| \cos \left( \frac{\theta_{j,1}}{2} \right)  \right )\hat{a}_{j}^{\dagger}.
\label{eq:9}
\end{eqnarray}

Consequently, the contribution of $\hat{a}_{j+1}^{\dagger}$ can always be suppressed 
by choosing the appropriate angle, namely,

\begin{equation}
\tan \left( \frac{\theta_{j,1}}{2} \right) = \frac{|c_{j+1,1}|}{|c_{j,1}|}.
\label{eq:11}
\end{equation}

If the procedure is first utilized to suppress $\hat{a}_{N}^{\dagger}$, then
one can successively suppress the other ladder operators in decreasing order until
just $(\hat{a}_{1}^{\dagger})^{n_1}$ is left acting on $|0\rangle$. Here, this process 
is referred to as Folding. The inverse process, or {\it Unfolding}, is just a
way of getting the original state back

\begin{equation}
|\psi\rangle = \left ( \prod_{l=1}^{N} e^{ i \phi_{l,1} \hat{a}_l^{\dagger} \hat{a}_l } \prod_{j=1}^{N-1} e^{ i \theta_{j,1} \hat{J}_{j+1,j}^y}  \right ) \frac{\left ( \hat {a}_1^{\dagger} \right )^{n_1}}{\sqrt{n_1!}} | 0 \rangle.
\label{eq:12}
\end{equation}

Notice that now the order in which two-site transformations are applied matters.
The order of multiplication is assumed to be

\begin{equation}
\prod_{j=1}^{N-1} e^{ i \theta_{j,1} \hat{J}_{j+1,j}^y} = e^{ i \theta_{N-1,1} \hat{J}_{N,N-1}^y}\dots e^{ i \theta_{1,1} \hat{J}_{2,1}^y},
\label{eq:19}
\end{equation}

and analogously in subsequent expressions. The second subscript in the angles and
the coefficients makes reference to the only mode left
after Folding. In order to write (\ref{eq:12}) as a set of tensors, the state with $n_1$
bosons in the first place of the chain is written as MPS. This can be readily done
because the Schmidt vectors of such a state have a simple structure. Subsequently,
the tensorial representation is updated according to Eq. (\ref{eq:12}), following 
the protocols available for one- and two site operations \cite{Vidal}.

More complex situations take place when bosons are distributed over several 
SB states. As has been seen, an important class of these states can be generically 
represented as

\begin{equation}
\frac{1}{\sqrt{n_1!...n_{N'}!}} \left( \sum_{k=1}^N c_{k,N'} \hat{a}_{k}^{\dagger} \right)^{n_{N'}} \dots \left( \sum_{j=1}^N c_{j,1} \hat{a}_{j}^{\dagger} \right)^{n_1} |0\rangle,
\label{eq:13}
\end{equation}

together with 

\begin{equation}
\sum_{j=1}^N c_{j,l'} c_{j,l}^* = \delta_{l'}^l \text{ } (l,l'=1,\dots,N'), 
\label{eq:18}
\end{equation}

which requires $N' \leq N$. To fold (\ref{eq:13}), the first 
NDM is folded as shown for (\ref{eq:7}). This affect the coefficients
of the other NDMs but because the transformations are {\it linear in
the operators}, the new coefficients obey a relation like (\ref{eq:18}). 
As a result, after folding the first NDM, the coefficients of 
$\hat{a}_1^{\dagger}$ automatically vanish in the other NDMs. 
The process can then be applied again to fold the second NDM, but this
time it is more reasonable to fold until $\hat{a}_2^{\dagger}$ is left
alone, skipping the last folding operation, since $\hat{a}_1^{\dagger}$
is not present in the second NDM. In this way, folding the second
NDM does not unfold the first mode and $\hat{a}_2^{\dagger}$
disappears from the rest of NDMs. The procedure is repeated
in a similar way until the state is reduced to a simple product
of local Fock states. The original state (\ref{eq:13}) can therefore 
be recovered as

\begin{equation}
\left ( \prod_{k=N'}^1 \left ( \prod_{l=k}^{N} e^{ i \phi_{l,k} \hat{a}_l^{\dagger} \hat{a}_l } \prod_{j=k}^{N-1} e^{ i \theta_{j,k} \hat{J}_{j+1,j}^y}  \right ) \right ) \prod_{q=1}^{N'} \frac{\left ( \hat {a}_q^{\dagger} \right )^{n_q}}{\sqrt{n_q!}} | 0 \rangle,
\label{eq:14}
\end{equation}

which in turn can be numerically implemented in terms of MPS as
explained before.

\section{Applications} 

In order to test Unfolding in a controlled manner, a Hamiltonian
with a known analytical profile is brought up, namely

\begin{equation}
h_{j,k} = \delta_{k+1}^j + \delta_k^{j+1},
\label{eq:15}
\end{equation}

plus periodic boundary conditions, $h_{j,N+1}=h_{j,1}$ and $h_{N+1,k}=h_{1,k}$ 
($\hat{a}_{N+1}^{\dagger}=\hat{a}_1^{\dagger}$). As the spectrum of
this Hamiltonian is in general degenerate, the next reference eigensystem is chosen

\begin{equation}
\epsilon_l = 2 \cos \left( \frac{2 \pi l }{N}  \right), \text{ } \epsilon_{k,l} = \frac{ e^{2\pi k l i/N} }{\sqrt{N}}.
\label{eq:16}
\end{equation}

\begin{figure}
\begin{center}
\includegraphics[width=0.32\textwidth,angle=-90]{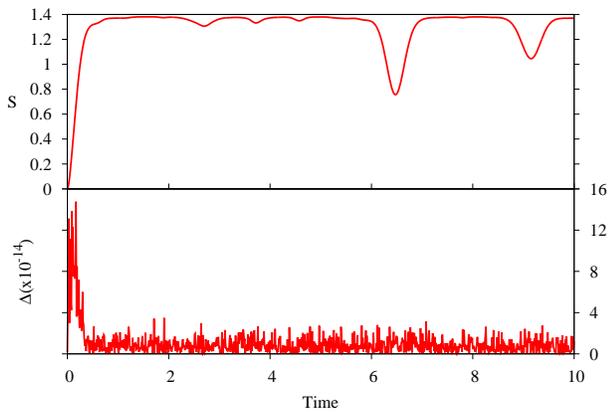}
\caption{Entropy between one site and the rest of the system (Top) and error $\Delta= 1 - |\langle \psi | \psi' \rangle |^2 $ (Bottom)
in a boson chain with $N=8$ and $M=8$ initialized with one particle at each site. The underlying Hamiltonian displays 
next-neighbor hopping (Eqs. (\ref{eq:1}) and (\ref{eq:15})) and the boundary conditions are periodic. 
The error determines the difference between 
the state found by standard diagonalization ($| \psi' \rangle$) and by Unfolding as explained in the text.} 
\label{fig1}
\end{center}
\end{figure}

The calculation consists in solving Eq. (\ref{eq:3}) and then
inserting the dynamical operators in Eq. (\ref{eq:4}), assuming that
at $t=0$ there is one boson at each site of the chain. The resulting 
state is then written as a tensor network using Unfolding.
To do this,  Eq. (\ref{eq:14}) is implemented as a numerical routine that integrates
the updating subroutine of the programs described in \cite{Reslen2}.
The obtained results are then compared against equivalent simulations
carried by diagonalization.
The lower panel of Fig. \ref{fig1} shows, as a function of time, 
the numerical error produced by Unfolding when compared to the 
standard method. Unless otherwise stated, it must be assumed that 
in the MPS computations $\chi$ is not bounded but dynamically 
determined by the updating routine as the simulation runs. In this 
way, all the elements of the MPS representation are retained.
As can be seen in Fig. \ref{fig1}, error is comparable to computer 
precision and it does not grow over long intervals. This because 
in Unfolding the state for a given time only depends on the 
initial condition and the solution of the equations of motion for 
the operators, which can be obtained with high accuracy for any $t$.
The upper panel of Fig. \ref{fig1} shows the single site entropy of 
the chain, calculated from

\begin{equation}
S = - \sum_\mu \left( \lambda_\mu^{[1]} \right)^2  \log \left( \lambda_\mu^{[1]} \right)^2.
\label{eq:20}
\end{equation}

$S$ measures the entanglement between one site and the rest of the
chain and can be easily computed from a MPS representation. 
It is known that the chain relaxes to a Gaussian state
with maximum entropy subject to fixed second moments \cite{Cramer}. 
As a result, the saturation of $S$ determines a time window along 
which the dynamics is relevant.  

\begin{figure}
\begin{center}
\includegraphics[width=0.32\textwidth,angle=-90]{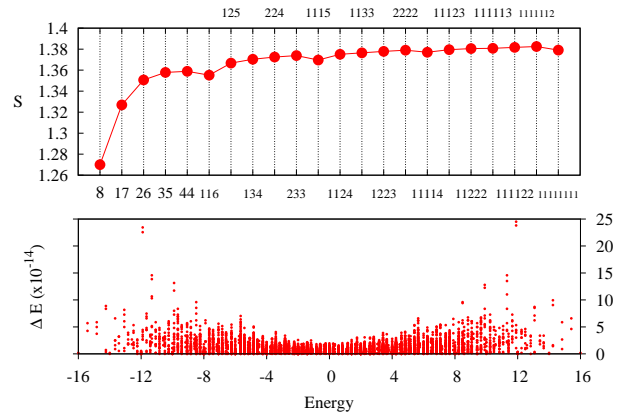}
\caption{Entropy between one site and the rest of the system (Top) and error 
$\Delta E = |E-E_{n_1,\dots,n_{N}}|$ (Bottom) of energy eigenstates of Hamiltonian (1) for $N=8$ and
$M=8$. The eigenstates are built as products of SB eigenmodes, as described by 
Eq. (\ref{eq:2}), and then converted to MPS in order
to find $S$ and $E$. Since $S$ depends only on the exponents
of the product, a many-particle state is represented by the number of bosons at each SB state, 
making no reference to which SB state the exponent actually apply. For instance, $17$ means
two SB states are involved, the first with $1$ boson and the second with $7$ bosons. Entropy is independent
on the specific choice of such SB states.} 
\label{fig2}
\end{center}
\end{figure}

Fig. \ref{fig2} shows $S$ for the eigenstates of $\hat{H}$
as well as the numerical error incurred by passing such eigenstates
to MPS using Unfolding. In this figure every state has been
represented only by the exponents that appear 
in Eq. (\ref{eq:2}). This can be done because a SB
eigenmode formed from (\ref{eq:16}) can be transformed into any other 
SB eigenmode of the same family using
only single-site unitary operations. Recall that invariance under local unitary
transformations is a property of entanglement. Fig. \ref{fig2} suggests that
eigenstates of $\hat{H}$ made of bosons distributed over many SB eigenstates contain more
entanglement than eigenstates with bosons arranged over few SB eigenstates. Nevertheless,
the eigenstate of $\hat{H}$ with all SB states occupied does not show maximum entanglement.

\begin{figure}
\begin{center}
\includegraphics[width=0.32\textwidth,angle=-90]{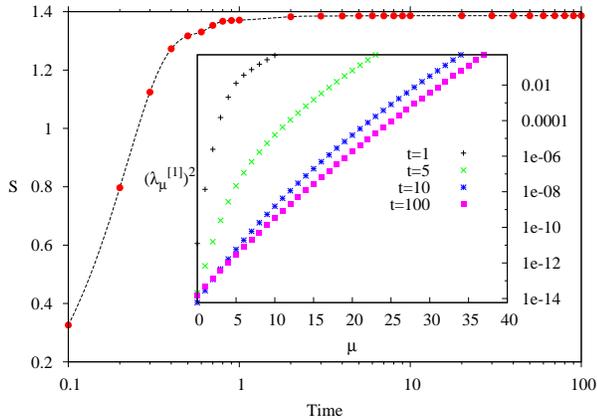}
\caption{Boson chain with $N=100$, $M=100$ and the same conditions as in Fig. \ref{fig1}.
In this example the size of the MPS representation was bounded by setting $\chi=50$. Inset. 
Eigenvalues of the single-site density matrix for different times. As the logarithmic
plot of the eigenvalue distribution becomes more linear, the state approaches a 
Gaussian state.} 
\label{fig3}
\end{center}
\end{figure}

The efficiency of Unfolding as a numerical method varies inversely to $\chi$.
In relation to this, the number of operations necessary to update the MPS representation
every time a unitary transformation is applied grows with the size of the 
local basis ($M+1$), but is attenuated by exploiting conservation
of number of particles. Moreover, from the arguments in \cite{Vidal} it 
follows that the number of operations required to update the state must 
grow as a polynomial
of $\chi$. This makes Unfolding suitable for systems with little
entanglement. However, because every time the state is computed 
only one round of unitary operations is invoked (Eq. \ref{eq:14}), 
Unfolding is different to methods where the calculation of the
state for a given time entails an integration of short
evolutions. The fact that in Unfolding error does not accumulate with time is
also an advantage, as well as the fact that specific choices of boundary
conditions or number of neighbors do not necessarily preclude the application
of the method. The key point is to put the state in
the form of Eq. (\ref{eq:13}). Likewise, the advantages of Unfolding
over diagonalization can be appreciated by noticing that while
the basis of $\hat{H}$ grows exponentially
with $N$, the bases of the matrices involved in the Unfolding calculation
grow linearly with $N$. In comparison to other methods that
could be applied in the same circumstances, Unfolding would be suitable
when a MPS description is preferred or when entanglement is small.

For states with large entanglement, Unfolding can be used to get 
an estimation. This is done by setting $\chi$ to a
numerically manageable value.  Fig. \ref{fig3} shows
some results obtained by fixing $\chi$ in a simulation of a 
relatively big chain. In spite of the approximation, the
relaxation profile shows good agreement with theoretical assessments 
reported in \cite{Cramer}.

\section{Discussion and outline}

Although Unfolding has been presented in the context of a specific
class of initial states, 
it appears the same strategies can in principle be applied whenever
the state is in general given by

\begin{equation}
f \left ( \sum_{k=1}^N c_{k,N'} \hat{a}_{k}^{\dagger}, \dots, \sum_{j=1}^N c_{j,1} \hat{a}_{j}^{\dagger} \right ) | 0 \rangle,
\label{eq:50}
\end{equation}
   
as long as $f$ could be expanded in Taylor series. Furthermore, coherent
states like

\begin{equation}
e^{\sum_j \alpha_j \hat{a}_j^{\dagger} - \alpha_j^* \hat{a}_j }  | 0 \rangle,
\label{eq:51}
\end{equation}

exhibit some compatibility with Unfolding too. In these cases, Folding
would reduce the state to a function of ladder operators
acting on $|0\rangle$. The success of the method would then
depend on the possibility of writing such a reduced state
in MPS terms without much effort. The translated state can then
be used as the initial condition in a simulation effectuated by,
for instance, TEBD. 

As commutativity of NDMs (Eq. (\ref{eq:18})) is assumed in
Unfolding, non-commuting NDMs can be treated by adding
modes that correct this anomaly. As an example, consider
the state

\begin{equation}
\left ( \sum_{j=1}^2 c_{j,1} \hat{a}_j^{\dagger} \right ) \left ( \sum_{k=1}^2 c_{k,2} \hat{a}_k^{\dagger} \right ) | 0 \rangle, \text{ } \sum_{j=1}^2 c_{j,1} c_{j,2}^* \ne 0.
\label{eq:52}
\end{equation}

A third mode can be introduced so that 

\begin{equation}
\left ( \sum_{j=1}^3 c_{j,1} \hat{a}_j^{\dagger} \right ) \left ( \sum_{k=1}^3 c_{k,2} \hat{a}_k^{\dagger} \right ) | 0 \rangle, \text{ } \sum_{j=1}^3 c_{j,1} c_{j,2}^* = 0.
\label{eq:52}
\end{equation}

Up to a normalization constant, the new state can be folded 
as shown above. Once the transformation to MPS has been carried, the
coefficients related to the extra mode can be dropped.
This approach is resembling of density matrix purification.
On the other hand, one way of taking
interaction effects into account is to apply perturbation
theory, treating non-linear terms as perturbations.
This would result especially effective when the non-linearity
is local, because the MPS description is appropriate to find local mean values.
Another way is to mimic the interaction using a mean-field  
approach. This could be realized by using the solution of the non-linear
Gross-Pitaevskii equation as the coefficients of Eq. (\ref{eq:7}). 
One can also think of using Eq. (\ref{eq:14}) as a variational ansatz,
similar to the Gutzwiller ansatz.

An alternative method has been proposed in the context of linear bosonic
systems to compute physical quantum states in MPS form. The technique
has been used to simulate an understood model and the results have
been compared against both data produced by diagonalization and
theoretical studies. Agreement has been satisfactory in 
every case. Aspects related to the suitability and scope 
of the technique have been analyzed and complementary observations
have been made.


\begin{thebibliography}{}

%\bibitem{Reslen1} J. Reslen, {\it Braz. J. Phys.} {\bf 0}, 0 (2013).

\bibitem{Ulrich} U. Schollw\"ock, Annals of Phys. {\bf 326} 96 (2011).

\bibitem{Cirac1} M.C. Ba\~nuls, M.B. Hastings, F. Verstraete and J.I. Cirac, Phys. Rev. Lett. {\bf 102} 240603 (2009), N. Schuch, M.M. Wolf and J.I. Cirac, arXiv:1201.3945. %Matrix product states for dynamical simulation of infinite chains. - names its technique as folding, Gaussian MPS.

\bibitem{Evenbly} G. Evenbly and G. Vidal, New J. Phys. {\bf 12} 025007 (2010), G. Evenbly and G. Vidal, arXiv:1210:1895. % Entanglement renormalization in free-bosonic systems: real-space versus momentum-space renomalization group transforms, A class of highly entangled many-body states that can be efficiently simulated.

\bibitem{Danshita} R.V. Mishmash, I. Danshita, C.W. Clark, L.D. Carr, Phys. Rev. A {\bf 80} 053612 (2009). %Quantum many-body dynamics of dark solitons in optical lattices

\bibitem{Muth} D. Muth and M. Fleischhauer, Phys. Rev. Lett. {\bf 105} 150403 (2010). % Dynamics of pair correlations in the attractive Lieb-Liniger gas

\bibitem{Hu} A. Hu, L. Mathey, C.J. Williams and C.W. Clark, Phys. Rev. A {\bf 81} 063602 (2010). % Noise correlations of one-dimensional Bose mixtures in optical lattices

\bibitem{Lacki} M. Lacki, D. Delande and J. Zakrzewski, Phys. Rev. A {\bf 86} 013602 (2012). % Numerical computation of dynamically important excited states of many-body systems

\bibitem{Reslen2} J. Reslen and S. Bose, Phys. Rev. A {\bf 80}, 012330 (2009), J. Reslen, arXiv:1002.4001. 

\bibitem{Prior} S.R. Clark, J. Prior, M.J. Hartmann, D. Jaksch and M.B. Plenio, New J. Phys. {\bf 12} 025005 (2010). %Exact matrix product solutions in the Heisenberg picture of an open quantum spin chain

\bibitem{Buzek} M.S. Kim, W. Son, V. Buzek and P.L. Knight, Phys. Rev. A {\bf 65} 032323 (2002).

\bibitem{Vidal} G. Vidal, Phys. Rev. Lett. {\bf 91}, 147902 (2003).

\bibitem{Cramer} M. Cramer, C.M. Dawson, J. Eisert and T.J. Osborne, Phys. Rev. Lett. {\bf 100}, 030602 (2008).

\end{thebibliography}
\end{document}